\documentclass[fleqn,twoside]{article}
\usepackage{espcrc2}
\usepackage{graphicx}

\newcommand{\be}{\begin{equation}}
\newcommand{\ee}{\end{equation}}
\newcommand{\bea}{\begin{eqnarray}}
\newcommand{\eea}{\end{eqnarray}}
\newcommand{\M}{{\cal M}}
\newcommand{\nn}{\nonumber}

\title{Fermion Propagator in Quenched QED3 in the Light of the 
Landau-Khalatnikov-Fradkin Tranformation}

\author{A. Bashir\address[AB]{Instituto de F{\'\i}sica y Matem\'aticas,
Universidad Michoacana de San Nicol\'as de Hidalgo,\\ Apartado Postal
2-82, Morelia, Michoac\'an 58040, M\'exico.},
        A. Raya\address{Facultad de Ciencias, Universidad de Colima,\\
Bernal D\'{\i}az del Castillo \#340, Col. Villa San Sebasti\'an, Colima,
Colima 28045, M\'exico.}
}
       
\begin{document}

\begin{abstract}

We study the gauge dependence of the fermion propagator in quenched QED3, with 
and without dynamical symmetry breaking, in the light of its 
Landau-Khalatnikov-Fradkin transformation (LKFT). In the former case, 
starting with the massive bare propagator in the Landau gauge, we obtain 
non perturbative propagator in an arbitrary covariant gauge. Carrying out a
perturbative expansion of this result, it yields correct wavefunction 
renormalization and the mass function up to the terms independent of the 
gauge parameter. Also, we obtain valuable information for the higher order 
perturbative expansion of the propagator.  As for the case of dynamical 
chiral symmetry breaking, we start by approximating the numerical solution 
in Landau gauge in the rainbow approximation in terms of analytic functions. 
We then use LKFT to obtain the dynamically generated fermion propagator in an 
arbitrary covariant gauge. We find that the results obtained have all the
required qualitative features. We also go beyond the rainbow
and encounter similar desirable qualitative features.
\vspace{1pc}
\end{abstract}

\maketitle

\section{Introduction}

Quantum electrodynamics in a plane (QED3) continues to attract attention
both in the field of super-conductivity, {e.g.},~\cite{supercon}, where it 
has been used in the study of high $T_c$ super-conductors, as well as in the 
realm of dynamical mass generation (DMG) where the numerical findings on the 
lattice and the results obtained by employing Schwinger-Dyson equations 
(SDE),~\cite{differences}, are yet to arrive at a final consensus. 

In the context of the SDE, the full fermion propagator (FP) is related to the 
full photon propagator (PP) and the full fermion-boson vertex (FBV), which 
obey their own SDEs. The study of the DMG is related to the knowledge of the 
FP. Quenched approximation consists of neglecting fermion loops. 
Within the realm of this approximation, we still need an \emph{ansatz} for 
the FBV to study the FP. In the rainbow approximation, the FP equation 
decouples also from that of the FBV. The corresponding SDE gets even more simplified and one can 
extract the key features of a dynamically generated FP. However, it is known 
that the bare vertex violates gauge invariance. In order to improve our 
approximations, we must impose gauge invariance constraints on the FBV, 
like Ward-Green-Takahashi identity (WGTI)~\cite{WGTI}, the Nielsen identities 
(NI)~\cite{Nielsen} and the Landau-Khalatnikov-Fradkin transformations (LKFT). 
Perturbation theory is the only known scheme where all of these identities 
and the gauge independence of physical observables can be achieved at every 
order of approximation. Therefore, we probably stand our best chance to 
achive these 
features also non perturbatively  if we construct a  FBV which  reduces to its 
perturbative counterpart in the weak coupling regime.
This fact was recently exploited for the construction of the FBV in quenched 
QED3 by the authors \cite{BR1}. 
A test of such constructions is to study the resulting SDE
for various gauges, going in small steps of the gauge parameter away from the 
Landau gauge. It is prohibitively difficult to be able to compute the 
result for an arbitrarily large value of the gauge parameter especially if a 
sophisticated form of the three point interaction is taken into 
account,~\cite{BHR}. Fortunately, one of the gauge invariance constraints,
namely the LKFT, can help us to circumvent this problem.

The LKFT of the Green functions describe the specific manner in which these 
functions transform under a variation of gauge. These transformations are 
non perturbative in nature,  and they are better described in coordinate space.
The LKFT provide us with a mechanism to study the gauge dependence of the FP 
starting from its value in the Landau gauge. Both in the perturbative or 
non perturbative calculations, the knowledge of the gauge dependence of the 
Green functions is useful and the LKFT can help us to achieve this goal.

This work is organized as follows~: In next section we recall how to perform 
the LKFT for the FP. Sect. 3 is devoted to the non perturbative FP obtained 
fom the LKFT of the lowest order FP. We find valuable information in the 
perturbative expansion of this result. Sect. 4 is devoted to the LKFT of a 
dynamically generated FP obtained in Landau gauge with the bare vertex.
We also extend these studies to the case of 
a solution for the SDE considering the full FBV (hereafter we will use the 
notation FV for the full vertex and BV for the bare one). 
Sect. 5 contains the numerical findings.
Finally in Sect. 6 we present our conclusions and outlook.

\section{LKFT~: The Procedure}

We start by putting forward the definitions and notations we shall use along 
the way~\cite{LKF1,LKFdin}. We write out the FP in Euclidean momentum and 
coordinate spaces, respectively, in its most general form as~:
\bea
S(p;\xi)
&=& \frac{F(p;\xi)}{i {\not \! p}-\M(p;\xi)}\;,\label{fpropmoment} \\
S(x;\xi)&=&{\not \! x}X(x;\xi)+ Y(x;\xi) \label{fpropcoord} \;.
\eea
$F(p;\xi)$ is generally referred to as the wavefunction renormalization, 
whereas $\M(p;\xi)$ as the mass function. The above  expressions are related 
through the following Fourier transformations
\bea
S(p;\xi)&=&\int d^3x e^{i p\cdot x} S(x;\xi)\;, \label{fourierp}\\
S(x;\xi)&=&\int \frac{d^3p}{(2\pi)^3} e^{-ip\cdot x}S(p;\xi)  \;.   
\label{fourierx}
\eea
The LKFT relating the coordinate space FP in  the Landau gauge to the one in 
an arbitrary covariant gauge reads
\be
S(x;\xi)=S(x;0)e^{-a x} \;, \label{LKFTp}
\ee
where $a=\alpha \xi/2$. The way we proceed is as follows.  We start with 
a FP given in the Landau gauge and Fourier transform it to the coordinate 
space. We then apply the LKFT law. Fourier transform of this result back to 
the momentum space yields the FP in an arbitrary covariant gauge.

\section{LKFT of the Tree Level FP}

An illustrative example to understand the usage and implications of
the LKFT is to start from the lowest order fermion propagator in the
Landau gauge.
\be
 F(p;0)=1 \; \; \mbox{and} \; \; \M(p;0)=m \;.  \label{lowestFM}
\ee
After performing the the LKFT through the procedure outlined 
before~\cite{LKF1}, we find that~:
\bea
&& \hspace{-6mm}  F(p;\xi)=-\frac{a}{p}
\arctan{\left[\frac{p}{m+a} \right]}
+ \frac{8p(p^2+a^2)}{ \phi(p;\xi)  }\nn\\
&& \hspace{5mm} -
\frac{8a (p^2+a(m+a))}{\phi(p;\xi)}\arctan{\left[\frac{p}{m+a}\right]} \;,  \label{LKFF3} \\
&& \hspace{-6mm} \M(p;\xi)=\frac{8p^3 m}{\phi(p;\xi) } \;,   \label{LKFM3}
\eea
where
\bea
&& \hspace{-6mm} \phi(p;\xi) = 8p(p^2+a(m+a))\nn\\
&& \hspace{5mm}
 -
8a(p^2+(m+a)^2)\arctan{\left[\frac{p}{m+a}\right]} \;.
\eea
In the weak coupling, we can expand out Eqs.~(\ref{LKFF3},\ref{LKFM3})
in powers of $\alpha$. To ${\cal O}(\alpha)$, we find
\bea
&& \hspace{-8mm} F(p;\xi)= 1\!+\!\frac{\alpha\xi}{2p^2}\left[(m^2\!-\!p^2) 
I(p) \!-\! m p^2\right] ,   \\
&& \hspace{-8mm} \M(p;\xi)= m \hspace{-1.2mm} \left[ 1
\!+\!\frac{\alpha\xi }{2p^2}\left\{ (m^2\!+\!p^2)  I(p) \!-\! m p^2 \right\}  
\right],
\label{LKFM1lazo}
\eea
where
\be
I(p)=\frac{1}{p}\arctan{\left[ \frac{p}{m}\right]}\;.
\ee
Let us compare these results with the one loop results obtained 
in~\cite{BR1}~:
\bea
&& \hspace{-6mm} F_{1{\rm l}}(p;\xi)=
1\!+\!\frac{\alpha\xi}{2p^2}\left[(m^2\!-\!p^2)I(p) \!-\! m p^2\right] ,   
\label{QED3F1lazo} \\
&& \hspace{-6mm} \M_{1{\rm l}}(p;\xi)=m\nn\\
&&\hspace{-6mm}\times
\left[ 1\!+\!\frac{\alpha}{2p^2}
\left\{ [\xi (m^2\!+\!p^2) \!+\! 4 p^2]I(p) \!-\!\xi m p^2 \right\} \right] .    \label{QED3M1lazo}
\eea
The subscript ${\rm 1l}$ indicates that the quantities evaluated are
at the one loop level.
We see that the results obtained from the LKFT of a tree level FP 
are in accordance with the one loop FP upto a term which does not vanish in 
Landau gauge. Therefore, the knowledge of the lowest order FP, in conjunction 
with the LKFT, is sufficient to know all the
gauge dependent pieces of the FP at the one loop level.
The structure of the LKFT is such that the FP of order 
${\cal O}(\alpha^n)$ in the Landau gauge 
 fixes the coeficients of all the terms of the form
$
\alpha^{i+n}\xi^i$ for $i=0,1,\ldots$
in addition to the ones of higher power in $\xi$ at a given order in 
$\alpha$.

\section{DMG in Rainbow Approximation}
In the rainbow approximation, one can write the SDE for 
the FP as~:
\bea
\frac{1}{F(p;\xi)}&=&1-\frac{\alpha\xi}{\pi p^2} \int_0^\infty dk k^2 
K_F(k;\xi) \nn\\
&&\times\,
\left[ \;
1-\frac{k^2+p^2}{2kp}\ln{\left|\frac{k+p}{k-p}\right|} \; \right]\;,
\label{eqforF}   \\
\frac{\M(p;\xi)}{F(p;\xi)}&=&\frac{\alpha(\xi+2)}{\pi p} \int_0^\infty 
dk k K_M(k;\xi)\nn\\
&&\times\ln{\left|\frac{k+p}{k-p}\right|} \;,  \label{eqforM}
\eea
with
\be
K_F(p;\xi)=\frac{F(p;\xi)}{p^2+{\cal M}^2(p;\xi)}=\frac{K_M(p;\xi)}{\M(p;\xi)}\;.
\ee
 Owing to the fact that in the Landau gauge, $F(p;0)=1$, it has 
long served as a favourite gauge for the numerical study of these equations.
In this gauge, one only has to solve the following equation~:
\be
{\cal M}(p;0)=\frac{2 \alpha }{\pi p}\int_0^\infty dk k
K_M(k;0)\ln{\left|\frac{k+p}{k-p}\right|}\;.\label{MFint}
\label{LGMF}
\ee
The corresponding numerical solution has the following key features~: It 
behaves like a constant for low-$p$ and falls as $1/p^2$ for large 
momentum~\cite{BHR}. We can therefore approximate this behavior with 
simple analytical functions and perform the LKFT exercise to see if 
those key features remain intact or get modified.
We can approximate the numerical solution by the following function~: 
\be
{\cal M}(p;0)=\frac{M_0 \, m_0^2}{p^2+m_0^2}\;.\label{param}
\ee

\begin{figure}[t!] 
\vspace{-3.2cm}
{\centering
\resizebox*{0.4\textwidth}{0.4\textheight}
{\includegraphics{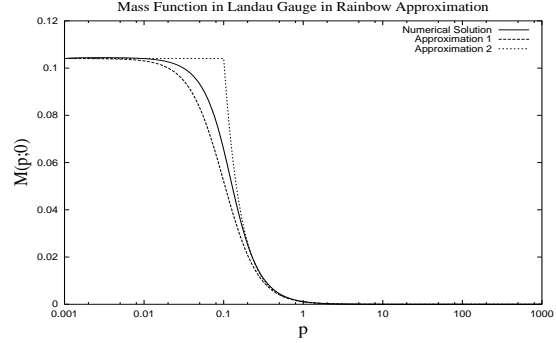}}
\par}
\vspace{-40pt}
\caption{The mass function in the Landau gauge for the 
bare vertex. Approximations proposed in Eq.~(\ref{param}) and 
Eq.~(\ref{param1}) are also shown.}
\label{fig1}
\end{figure}

\noindent
With this approximation, we find that the LKFT excercise is still not trivial 
to perform. Therefore, we need a further simplification.  We can write 
another approximation to the mass function as follows~:
\be
{\cal M}(p;0)= M_0 \left[ \theta(m_0-p) + \frac{m_0^2}{p^2} \theta(p-m_0) 
\right]
\;. \label{param1}
\ee
As shown in Fig.~(\ref{fig1}), Eqs.~(\ref{param},\ref{param1}) provide a 
good  approximation. We are now in a position to use LKFT to find the fermion
propagator in an arbitrary covariant gauge.

\subsection{LKFT for the FP in Rainbow Approximation}

The LKFT exercise for the FP in rainbow approximation can be performed,
\cite{LKFdin}, to analyse the dynamically generated behaviour of the FP
in the  asymptotic limits of momenta, i.e., when $p>>m_0$ and $m_0>>p$. 
In the large-$p$  limit, the mass function and 
the wavefunction renormalization have been found to have the following form~:
\bea
{\cal M}(p;\xi)&=&  
\frac{C_3(\xi)}{p^2}   +{\cal O}\left( \frac{1}{p^3}\right) \;,\\
F(p;\xi) &=&1+{\cal O}\left( \frac{1}{p}\right) \;,
\eea
where $C_3(\xi)$ is given in Ref.~\cite{LKFdin}.
An analogous analysis for the low-$p$ regime yields~:
\bea
{\cal M}(p;\xi)&=& \frac{C_1(\xi)}{C_2(\xi)} + {\cal O}(p^2) \;, \\
F(p;\xi)&=&-\frac{C_1^2(\xi)}{C_2(\xi)}-C_2(\xi) p^2  
+ {\cal O}(p^4)\;.
\eea
$C_1(\xi)$ and $C_2(\xi)$ are also given in Ref.~\cite{LKFdin}. Expectedly, 
$\M$ is flat for small values of $p$, and it falls off as $1/p^2$ for its 
large values. On the other hand, $F$ is also a constant for small values of 
$p$. For the large values, it approaches 1.  Thus the $p$-dependence of the 
dynamically generated FP for the small and large values of the momentum 
continues to have the same qualitative features in an arbitrary covariant 
gauge as the ones in the Landau gauge and in its neighbourhood,~\cite{BHR}.

\subsection{DMG with FV}

As we mentioned early, a sophisticated \emph{ansatz} for the FV must be used 
into the SDE instead of considering the BV.  The latest in a  series of 
proposals 
for the FV in QED3 is the one suggested  in~\cite{BR1}. However, its 
employment to solve the SDE for the FP is a formidable task even in the 
Landau gauge. Let us concentrate on all vertices whose transverse part 
vanishes in Landau gauge, e.g.,~\cite{BC}. In this case, the numerical 
behaviour of $F$ modifies to the one shown in Fig.~(\ref{fig2}). It behaves 
like a constant (different from unity) for low momentum, and tends to one 
as $p\to\infty$. Therefore, in addition to using the approximation in 
Eq.~(\ref{param1}) for $\M$ (with $M_0\to M_{0F}$ and $m_{0}\to m_{0F}$ 
because the solutions in the Landau gauge for the BV and the FV are not 
identical), we can use the following simple form for 
$F(p;0)$, Fig.~(\ref{fig2})~:

\begin{figure}[t!] 
\vspace{-3.2cm}
{\centering
\resizebox*{0.4\textwidth}{0.4\textheight}
{\includegraphics{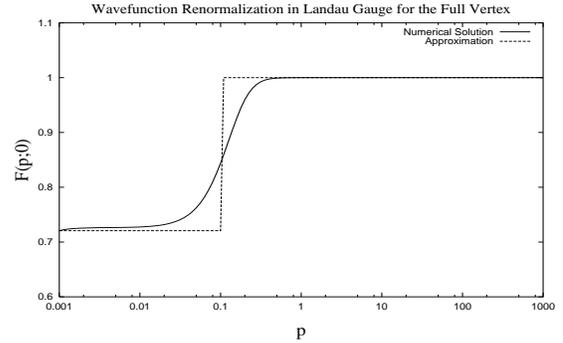}}
\par}
\vspace{-40pt}
\caption{$F(p;0)$ for the FV. Approximation proposed in 
Eq.~(\ref{fullparam}) is also shown.}
\label{fig2}
\end{figure}

\vspace{-3mm}
\bea
F(p,0) &=& F_{0F} \theta(m_{0F}-p) +\theta(p-m_{0F})  \label{fullparam}  \;.
\eea
The LKFT exercise is rather straightforward~\cite{LKFdin}, since it involves 
only slight modifications to the coefficients $C_1(\xi)$, $C_2(\xi)$ and 
$C_3(\xi)$ and the large- and low-$p$ behaviour of the FP remains 
essentially
unchanged. $\M(p;\xi)$ behaves like a constant for low-$p$ and falls as 
$1/p^2$ in an aritrary covariant gauge. $F(p;\xi)$ is a constant for 
low-$p$ and goes to 1 as $p$ reaches a large number.  These  modifications 
are taken into account in the modified expressions
for $C_{1F}(\xi)$, $C_{2F}(\xi)$ and $C_{3F}(\xi)$ given also in 
Ref.~\cite{LKFdin}.

\section{Numerical Findings}

\begin{figure}[t!] 
\vspace{-3.5cm}
{\centering
\resizebox*{0.4\textwidth}{0.4\textheight}
{\includegraphics{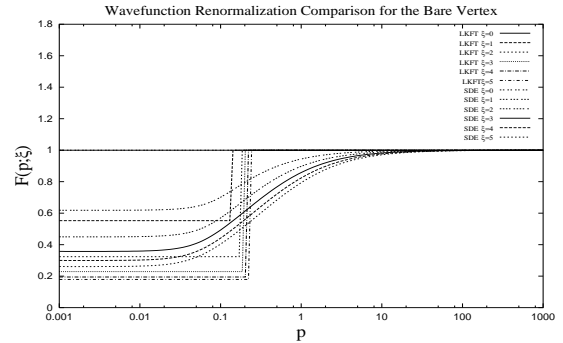}}
\par}
\vspace{-40pt}
\caption{$F(p;\xi)$ for the BV employing LKFT.
For a comparison, we also plot the results obtained by directly solving
SDE.}
\label{fig3}
\end{figure}

\begin{figure}[t!] 
\vspace{-3.5cm}
{\centering
\resizebox*{0.4\textwidth}{0.4\textheight}
{\includegraphics{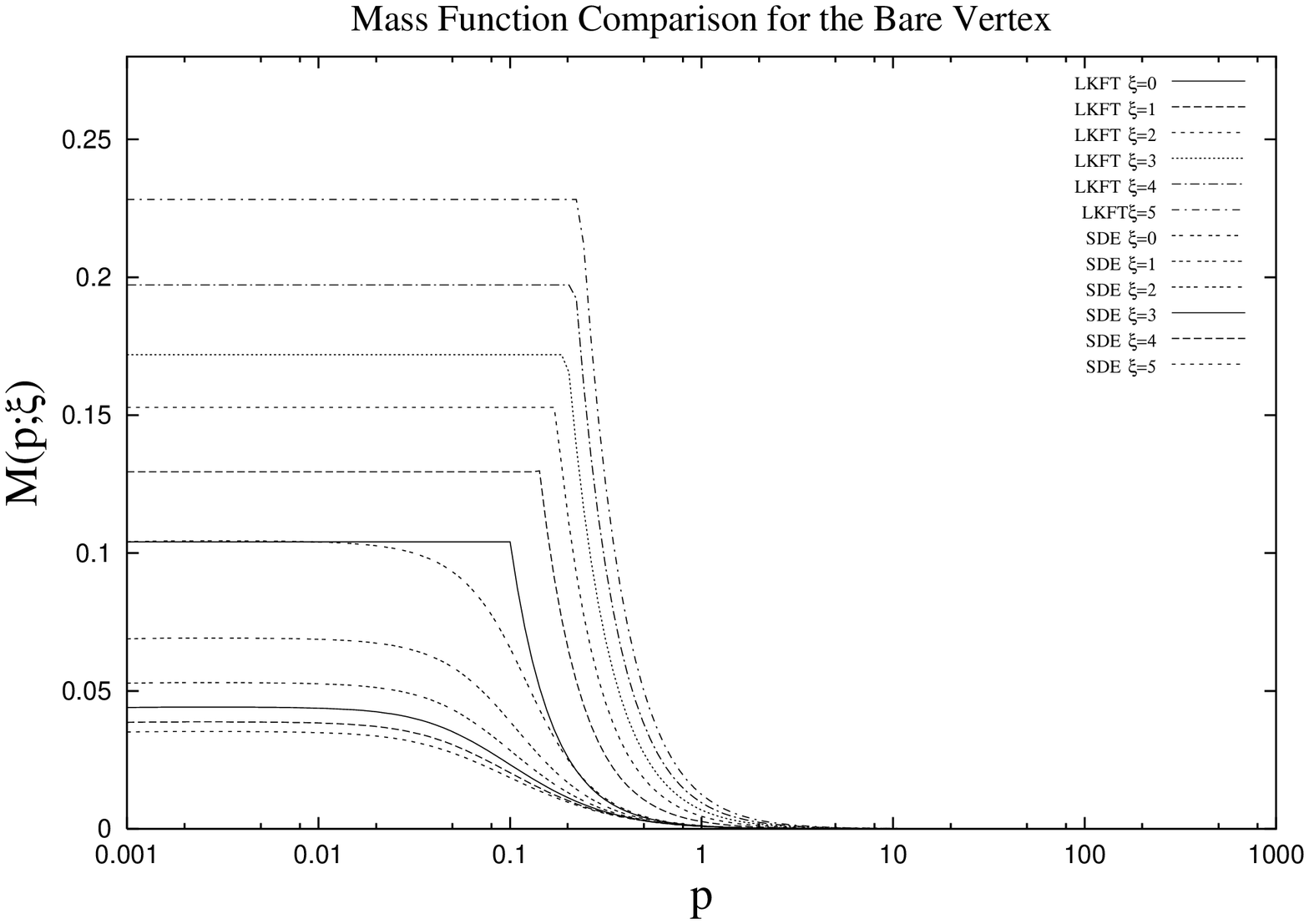}}
\par}
\vspace{-40pt}
\caption{$\M(p;\xi)$ for the BV employing LKFT.
For a comparison, we also plot the results obtained by directly solving
SDE.}
\label{fig4}
\end{figure}

\begin{figure}[t!] 
\vspace{-3.5cm}
{\centering
\resizebox*{0.4\textwidth}{0.4\textheight}
{\includegraphics{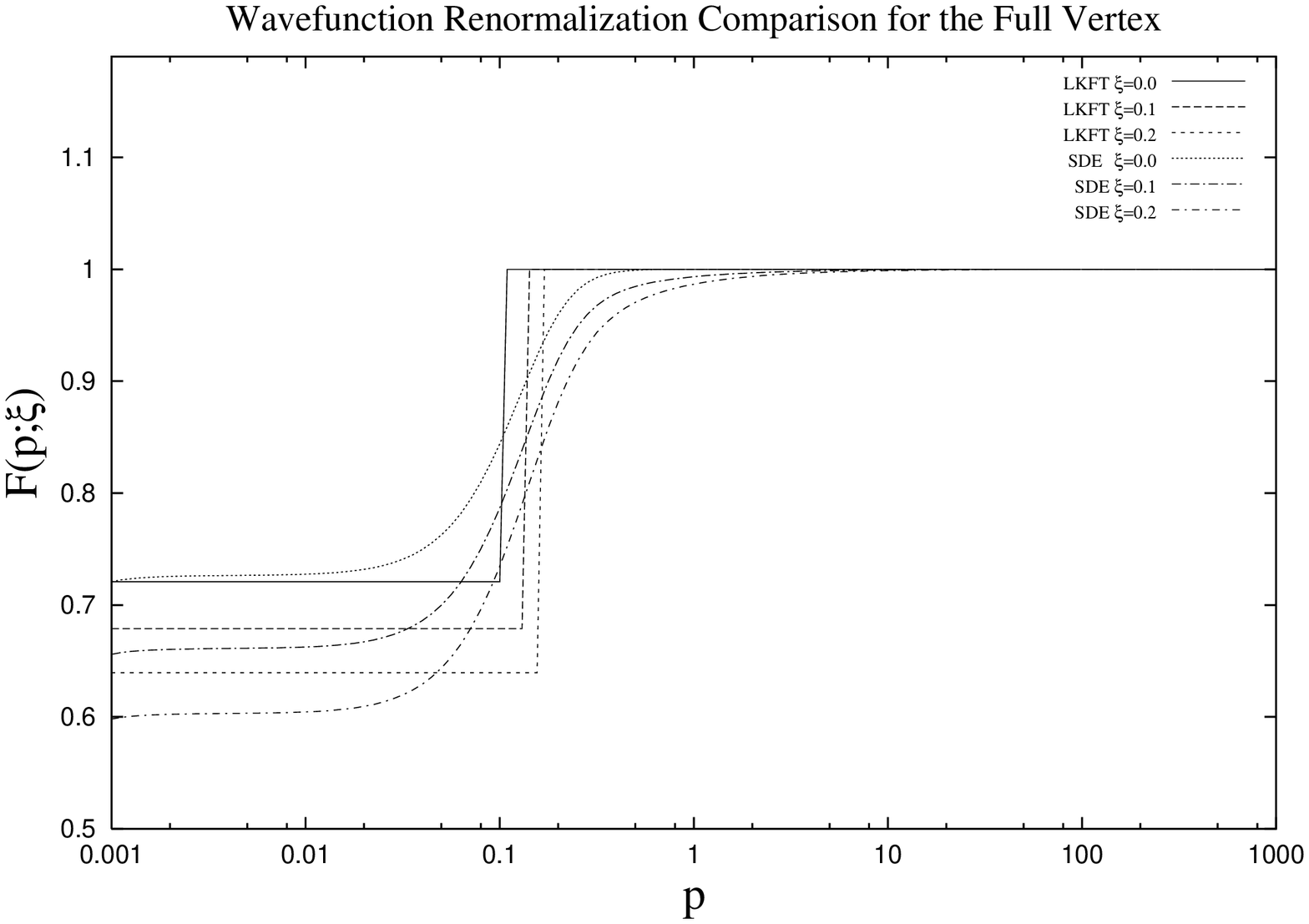}}
\par}
\vspace{-40pt}
\caption{$F(p;\xi)$ for the FV employing LKFT.
For a comparison, we also plot the results obtained by directly solving
SDE.}
\label{fig6}
\end{figure}

\begin{figure}[t!] 
\vspace{-3.5cm}
{\centering
\resizebox*{0.4\textwidth}{0.4\textheight}
{\includegraphics{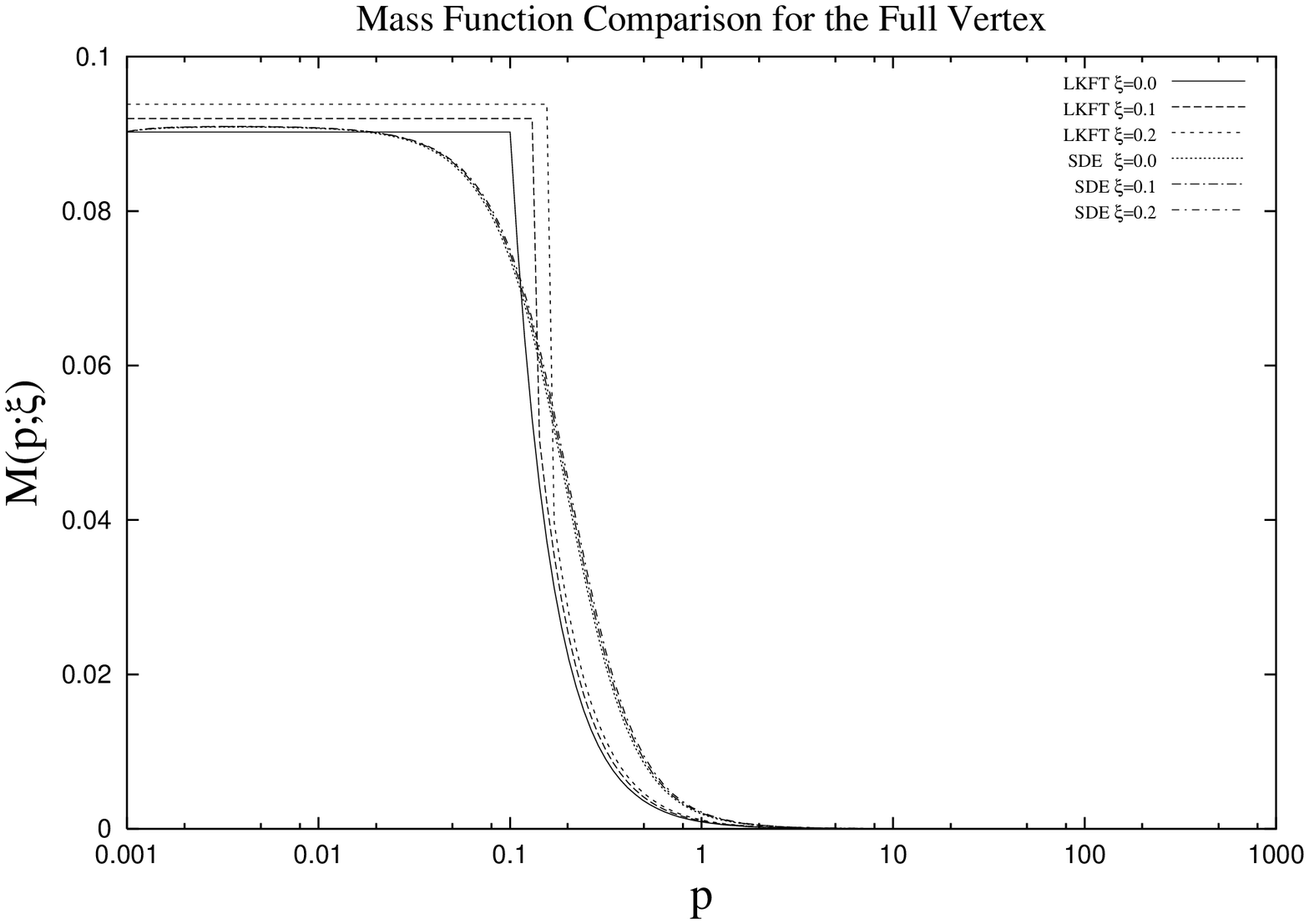}}
\par}
\vspace{-40pt}
\caption{$\M(p;\xi)$ for the FV employing LKFT.
For a comparison, we also plot the results obtained by directly solving
SDE.}
\label{fig8}
\end{figure}

From the corresponding expressions in the low and large momentum regimes for 
$F$ and $\M$, we perform the following parametrisation for the FP in 
arbitrary gauge~:
\bea
\M (p;\xi)&=& M_{\xi(F)}
\left[ \theta_1
+ \frac{m_{\xi(F)}^2}{p^2} \theta_2  \right]  \;, \\ 
F(p;\xi) &=&  F_{\xi(F)} \theta_1
+  \theta_2  \;,
\eea
where $\theta_1=\theta(m_{\xi(F)}-p)$, $\theta_2=\theta(p-m_{\xi(F)})$.
Moreover,
\bea
 M_{\xi(F)} &=& \frac{C_{1(F)}(\xi)}{C_{2(F)}(\xi)},  
 \quad
M_{\xi(F)} m_{\xi(F)}^2 \,=\, C_{3(F)}(\xi), \nn\\
 F_{\xi(F)} &=& -\frac{C_{1(F)}^2(\xi)}{C_{2(F)}(\xi)} \;. 
\eea

\begin{itemize}

\item In Fig.~(\ref{fig3}), we have plotted $F(p;\xi)$ in several gauges. 
Comparing these graphs with the ones obtained by solving SDE with the BV 
 \emph{ansatz}, one sees that the difference is not enormous, reassuring the 
correctness of the method employed.
\item In Fig.~(\ref{fig4}), we have plotted $\M(p;\xi)$ in several gauges 
in order to compare the results of directly solving SDE against the 
ones obtained by employing the LKFT. 
\item In Fig.~(\ref{fig6}) we present a comparison of $F(p;\xi)$ from the 
LKFT exercise against the results of  directly solving SDE with the FV. The 
results are found to be in fairly good agreement.
\item In Fig.~(\ref{fig8}) we present a comparison of  $\M(p;\xi)$ from the 
LKFT exercise against the results of directly solving SDE with the FV. It is 
a bit hard to make the comparison since the SDE results are known only in a 
small region near the Landau gauge~\cite{BHR,Huet}. A huge advantage of using 
LKFT method is that the results for an arbitrary value of the covariant gauge 
parameter are readily available.
\end{itemize}

\section{Conclusions and Outlook}

Knowledge of the FP  at an arbitrary order in perturbation theory in an 
arbitrary covariant gauge is as useful as it is difficult.  
If we are dealing with perturbative calculations, the number of diagrams 
increases enormously as the order of approximation gets bigger.  Since the 
LKFT of a giver order FP already fixes some of the 
coeficients of the FP in its all orders expansion, it is worth making use of
it. If we are dealing with nonperturbative studies of SDE, where it is 
hard to obtain solutions for arbitrarly large values of 
the gauge parameter, LKFT comes to rescue once again. If we are 
capable of solving the SDE for the FP with a sophisticated \emph{ansatz} for 
the FBV in Landau gauge, its LKFT yields the
dynamically generated FP for an arbitrarly large value of the gauge 
parameter. We expect this procedure to be translated to more 
complicated theories, like QCD, with only slight modifications.

{\bf Aknowledgements} We acknowledge CIC, CONACyT and Alvarez Buylla 
for the financial grants. AR whishes to thank A. Williams, 
A. K{\i}z{\i}lers\"u 
and the CSSM staff for their hospitality and financial support during the 
QCD Downunder 2004 Workshop.

\end{document}